# MODELING STUDY OF LASER BEAM SCATTERING BY DEFECTS ON SEMICONDUCTOR WAFERS

Srikumar Sandeep[1] and Alexander Kokhanovsky[2]

[1]Ecole Polytechnique, Montreal, Canada
sandeepsrikumar2013@gmail.com

[2] University of Bremen, Germany
Alexander.Kokhanovsky@eumetsat.int

**ABSTRACT**

*Accurate modeling of light scattering from nanometer scale defects on Silicon wafers is critical for enabling increasingly shrinking semiconductor technology nodes of the future. Yet, such modeling of defect scattering remains unsolved since existing modeling techniques fail to account for complex defect and wafer geometries. Here, we present results of laser beam scattering from spherical and ellipsoidal particles located on the surface of a silicon wafer. A commercially available electromagnetic field solver (HFSS) was deployed on a multiprocessor cluster to obtain results with previously unknown accuracy down to light scattering intensity of -170 dB. We compute three dimensional scattering patterns of silicon nanospheres located on a semiconductor wafer for both perpendicular and parallel polarization and show the effect of sphere size on scattering. We further computer scattering patterns of nanometer scale ellipsoidal particles having different orientation angles and unveil the effects of ellipsoidal orientation on scattering.*

## 1. INTRODUCTION

Leading-edge semiconductor integrated circuits have entered the 14 nm technology node, there by continuing the trend of producing smaller, faster, cheaper, and energy efficient chips [1]. While this trend unquestionably stands out as one of the key enablers of the information age, it must be recognized that continuing this trend is becoming increasingly challenging because of the difficulty in fabricating increasingly complex nanometer scale layers reliably and in a cost effective manner. As feature sizes continue to decrease in semiconductor chips, low manufacturing yield becomes a critical problem. Yield losses are caused due to imperfections or defects that could occur in any of the hundreds of manufacturing process steps that transform a bare Silicon wafer into a modern integrated circuit. Stringent process controls are required to ensure that each process step creates a desired result within established tolerances. Optical and electron based wafer inspection tools play a crucial role in process control by providing much needed feedback on a previously performed process step. The feedback includes critical information on the number, sizes, shapes, and material properties of defects - much needed information that can be used to adjust process steps to eliminate those defects, especially the ones that are likely to affect yield. Finding and eliminating such yield-affecting defects as soon as they occur is imperative to limit yield loss, which otherwise may rapidly propagate to a large number of wafers in a high-speed modern semiconductor fab.

In the current 14 nm leading-edge technology node, reduced yield can often be correlated to the inability of current inspection tools to effectively detect defects that are smaller than 20 nm. Novel fabrication processes used in leading-edge nodes result in an increased defect density as well new types of defects. New materials such as strained silicon, high-k dielectrics, and III-V compounds improve transistor performance, but create new defects due to their atomic scale interactions with Silicon. Further, new three-dimensional (3D) transistor designs such as FinFETs and new 3D stacking methods such as Through Silicon Vias (TSVs) require complex fabrication processes that often lead to added defects [3]. Enabling these advances require the development of improved defect inspection systems with ultra-high sensitivity and throughput [1,2]. An understanding of light scattering by nanometer scale defects located on a semiconductor wafer is important in the design and development of such inspection systems [4]. Such an understanding requires accurate modeling of optical scattering by semiconductor defects on or under the wafer surface.

A survey of previous modeling work related to this problem can be found in [4,5]. Analytical methods such as Mie theory [6] can only be used for scattering due to isolated spherical particles. Mie theory does not take into account the effect of surface beneath the scattering particle. Previously published works include the application of approximate methods such as Discrete Dipole Approximation (DDA) [7, 8] for defects of radius larger than 100 nm when excited with plane wave of longer wavelengths (e.g. 532 nm). Furthermore, these analytical methods cannot be used for defects under the wafer or multilayer wafer substrates. Hence there is a need to perform accurate full wave simulations to incorporate complex scenarios that are realistic in leading-edge semiconductor fabrication. To the best of our knowledge, there is no published work on full wave analysis of Gaussian beam scattering due to particles/defects on semiconductor wafers. The multi-scale nature of the problem (i.e. a small defect on a large wafer), requires the use of numerical field solvers executed on a multiprocessor cluster to obtain accurate results. Our work stands out from the previously published results in terms of the use of cutting edge modeling resources, better accuracy attributed to full wave methods, Gaussian laser beam excitation instead of plane wave excitation, shorter wavelength of 193 nm, sub-40 nm defect sizes, and the flexibility to model different defect shapes, sizes, materials and multilayer substrates.

## 2. MODELING FRAMEWORK

In this work, we use a laser beam of 193 nm wavelength, a commonly used wavelength in leading semiconductor photolithography systems [9]. The refractive index of silicon at 193 nm is n = 0.883 + 2.778i [10]. The geometry of our light scattering problem is shown in Fig. 1. It consists of a cylindrical Silicon wafer, the top surface of which coincides with the xy plane. A Gaussian beam having a minimum beam width (at its waist) of 400 nm is incident on the silicon wafer at an angle of incidence equal to $30^o$ with respect to z-axis and the laser beam axis lies on the xz-plane (i.e. wave vector $\hat{k} = 0.5\hat{x} - 0.866\hat{z}$). The Gaussian beam focal point location coincides with the coordinate origin and the field amplitude at this point is 1 V/m. We consider two polarizations in this paper. For perpendicular polarization (s-polarization), the polarization vector is given by $\hat{e} = \hat{y}$. For parallel polarization (p-polarization), the polarization vector $\hat{e} = 0.866\hat{x} + 0.5\hat{z}$. The parameter of interest for optical detection is the far zone scattered power density. Throughout this paper, we have used $20\log_{10}|\overline{E}_{sc}(r = 1\text{ m}, \theta, \varphi)|$ to quantify scattered power density. $\overline{E}_{sc}(r = 1\text{ m}, \theta, \varphi)$ is the scattered electric field intensity as a function of polar angle θ and azimuth angle φ on a sphere of radius 1 meter. Since we are interested primarily in the scattered field above the wafer, θ varies from 0 to 90 degrees. However, φ ranges from 0 to 360 degrees. The goal of our framework is to accurately model scattered light intensity values at all θ and φ angles for a range of realistic defects on a semiconductor wafer surface. It should be noted that the defect sizes are in the nanometer range and hence all practical optical detectors would be in the far field scattering zone. Hence using a value of r = 1 m is just a reference value. Once the field at r = 1 m is known, the fields at any other value of r in the far zone can be calculated.

The diameter of silicon wafers used in modern semiconductor fabrication could be as large as 450 mm. Since this dimension is much larger than the wavelength under consideration, it is impractical to include the entire wafer in the simulation. Fortunately, since the Gaussian beam is spatially confined (unlike a plane wave), a smaller local area of the wafer effectively emulates large realistic wafer diameters. By using wafers with progressively increasing radii, it was found that the scattering pattern converges for radii greater than 1200 nm. Because of the finite spatial extent of the Gaussian beam, a further increase in the wafer radius did not result in a significant change in the scattering pattern. This was confirmed by plotting the magnitude of electric field on the $\varphi = 0^o$ plane vs. the wafer radii used in simulation. A 1600 nm radius wafer was used for the simulation results shown in this paper.

We used ANSYS HFSS for the simulation of the above mentioned model. A Perfectly Matched Layer (PML) was used to truncate the problem geometry. Gaussian beam excitation was used as the source. The Gaussian beam excitation option in HFSS allows the entry of arbitrary polarization, propagation vector, focal point location and beam width at focal point. Our framework is flexible enough to simulate a wide variety of incident angles, excitation wavelengths, beam widths, incident wave polarizations, defect shapes, defect dimensions, and defect and wafer surface materials.

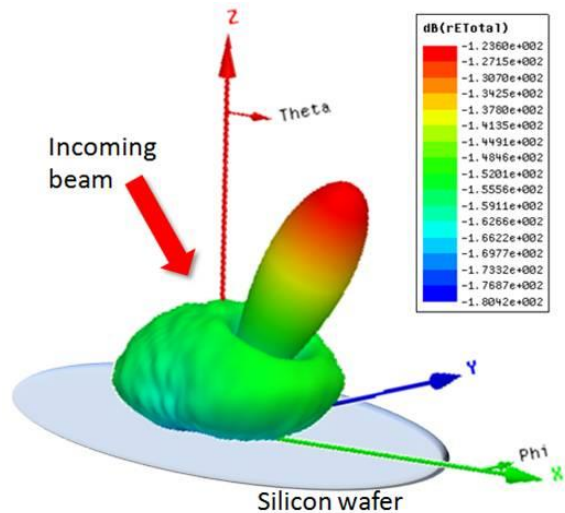

Figure 1. 3D scattering pattern: 40 nm radius sphere on wafer. The incident beam direction is given by $\hat{k} = 0.5\hat{x} - 0.866\hat{z}$.

## 3. RESULTS

We now show scattering patterns for spherical and ellipsoidal defects located on semiconductor wafers for s and p-polarizations. For each polarization, we start with a bare wafer (i.e. a wafer having no defects) to determine the noise floor of our model. This is followed by simulations of 20 nm, 30 nm and 40 nm radii silicon nanospheres located on the silicon wafer. Finally, we show simulations of a nanometer scale ellipsoidal particles located on a silicon wafer at $0^0$, $45^0$, $90^0$ and $135^0$ orientation angles.

### 3.1 Base Si wafer

When a bare wafer is illuminated with a Gaussian beam, the beam undergoes specular reflection. The specularly reflected beam propagates at $\theta = 30^0$ and $\varphi = 0^o$. In addition, weak scattering is observed at non-specular directions. Figure 2(a) shows the scattering pattern of the bare wafer for s-polarized illumination. This figure is a two-dimensional (2D) plot of $20\log_{10}|\overline{E}_{sc}(r = 1\text{ m}, \theta, \varphi)|$ as a function of $\theta$ and $\varphi$. The specularly reflected beam can be observed in the range $\theta \in [10^o, 50^o]$ and in the vicinity of $\varphi = 0^o$. For s-polarization, the peak of the specularly reflected beam is -122 dB, and the maximum value of the scattered field outside this specularly reflected beam is -170 dB. Figure 3(a) shows the scattering pattern of the bare wafer for p-polarized illumination. For the case of p-polarization, the peak of the specularly reflected beam is -127 dB, and the maximum value of the scattered field outside this specular beam is -171 dB. Accordingly, the computational noise floor of our model is -170 dB for s-polarization and -171 dB for p-polarization. Scattered light intensity from a defect needs to exceed this noise floor for the defect to be able to be detected reliably. It should be stated that in the 2D plots, the highest value of scattered field is truncated to -150 dB for visual clarity.

### 3.2 Silicon sphere on Silicon wafer: s-polarization

Figures 2 (b-d) show scattering patterns of silicon spheres with radii 20 nm, 30 nm and 40 nm located on the wafer surface for the case of s-polarized Gaussian beam excitation. Figure 1 is in fact a three-dimensional view of figure 2(d). In the 2D plots, the specular reflection is split into two halves along the $\varphi = 0^0, 360^0$ planes. This is because of the symmetry of the incident illumination beam with respect to xz plane. Consequently, the specularly reflected beam is also symmetric with respect to the xz plane. The maximum values of scattered radiation outside the specularly reflected beam are -164 dB, -159 dB, and -154 dB for the 20 nm, 30 nm and 40 nm spheres, respectively. It should be noted that for every 10 nm decrease in sphere radius, the maximum scattered field gets reduced by 5 dB. We find that scattering pattern is below the noise floor for spheres having a radius less than 10 nm with an s-polarized incident beam. The s-polarized incident beam induces an electric dipole oriented along the y direction. The radiation pattern of a dipole is dough-nut shaped with its null along its direction of polarization. These nulls can be seen in Figures 2(a-d), along $\varphi = 90^0, 270^0; \theta = 90^0$. The green ellipse in figure 2(d) denotes the null along $\varphi = 270^0$. As the radius of the sphere increases, the scattering pattern becomes asymmetric with more scattering along the forward direction (i.e. $\varphi = 0^0$).

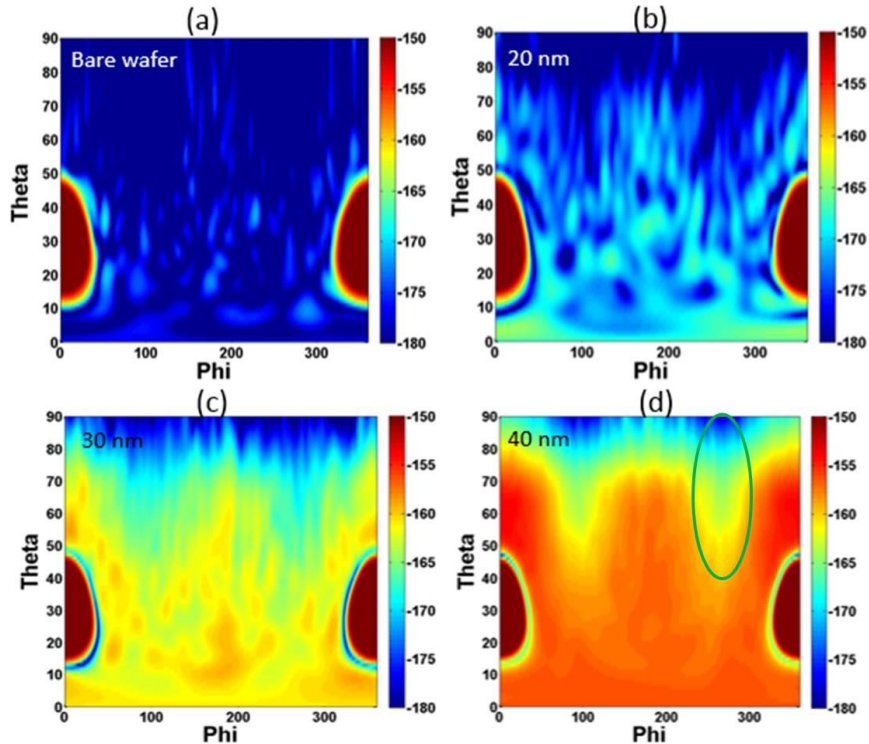

Figure 2. Light scattering by a) bare Si wafer, b) 20 nm sphere on a Si wafer, c) 30 nm sphere on a Si wafer and d) 40 nm sphere on a Si wafer, all of which are illuminated with a s-polarized Gaussian beam having a wavelength of 193 nm.

### 3.2 Silicon sphere on Silicon wafer: p-polarization

Figures 3 (b-d) show scattering pattern for silicon spheres with radii 20 nm, 30 nm and 40 nm located on the surface of a silicon wafer for the case of p-polarized Gaussian beam excitation. The maximum values of scattered radiation excluding the specularly reflected beam are given by -169 dB, -162 dB and -156 dB for the 20 nm, 30 nm and 40 nm spheres, respectively. We observe that the scattered energy in the case of p-polarization is less than that of s-polarization. We find that scattering pattern is below the noise floor for spheres having a radius less than 20 nm with a p-polarized incident beam. This is in contrast to 10 nm detection limit for s-polarization. Figure 4 shows a comparison of scattering patterns with s and p polarization incident beams for a 40 nm sphere on silicon wafer. Figure 4(a) illustrates this comparison in the $\varphi = 0^0$ plane, and Figure 4(b) illustrates this comparison in the $\varphi = 90^0$. Figure 4(a) shows that the scattered intensity for s-polarization is higher than the scattered intensity of p-polarization in the $\varphi = 0^0$ plane. On the other hand, Figure 4(b) shows that the scattered intensity for s-polarization rolls off with respect to azimuth angle at a faster rate than p-polarization. These observations can be explained by the rotation of the sphere scattering pattern when incident polarization is rotated from s to p. Because of the decreased scattered field intensity, s-polarization is preferred over p-polarization in industrial wafer inspection systems.

### 3.3 Silicon ellipsoid on Silicon wafer: s-polarization

We now show the scattering pattern of an ellipsoidal particle with axis lengths 140 nm x 60 nm x 60 nm located on a silicon wafer. The incident beam is s-polarized (i.e. $\hat{e} = \hat{y}$). Four different orientation angles 0◦, 45◦, 90◦ and 135◦ are used. Orientation angle denotes the angle between the major axis of the ellipsoid and the x-axis. For example, for the 45◦ case, the 140 nm length axis of ellipsoid makes an angle of 45◦ with x-axis. The major axis of the ellipsoid is parallel to the plane of the silicon wafer in all four orientation angle cases. Figure 5 shows the scattering pattern for the four cases. We observe that the scattering pattern for the ellipsoid with 90◦ orientation, shown in Figure 5(d), is similar to the scattering pattern of a 40 nm radius sphere shown in Figure 2(d). This is because of a couple of reasons: 1) volume of the 40 nm radius sphere is roughly equal to the volume of the ellipsoid; 2) major axis of ellipsoid with $90^0$ orientation is along the direction of incident beam polarization. From Figure 5, it can be observed that nulls occur along the direction of the ellipsoid major axis away from the specularly reflected lobe ($\varphi = 0^0, 360^0$). For instance, when the orientation angle is $45^0$ the major axis lies along $\varphi = 45^0, 225^0$ and the null occurs at $\varphi = 225^0$. A similar observation can be made in Figure 5(c) for the case of $135^0$ orientation angle.

## 4. CONCLUSIONS

This paper provides accurate full wave simulation results of laser beam scattering of spherical and ellipsoidal Silicon particles located on a cylindrical silicon wafer. The results presented in this paper is as close as to real world measurement scenario when compared to previous published results. The scattering pattern of a defect is found to have a close dependence on the shape of the defect. For s-polarization, the maximum scattered field outside the specularly reflected lobe reduces by 5 dB for every 10 nm reduction in sphere radius. Such a dependence can be used to detect the size of the defect. For p-polarization, the scattered energy is less than that of s-polarization. For the case of ellipsoidal defects, the scattering pattern exhibits a rotation with respect to the orientation of the major axis of the ellipsoid. Wafer surface roughness can have a profound effect on the scattering pattern and ultimately determines the smallest defect that can be detected. Although we ignore roughness in this paper, defect scattering pattern by itself could help us build more sensitive instruments. We believe the results in this work will be invaluable to wafer inspection technology companies, who can use these data to design novel instruments as well validate their numerical models [11]. Future work will include scattering study of more complicated defects such as Crystal Originated Pits (COPs) and different materials.

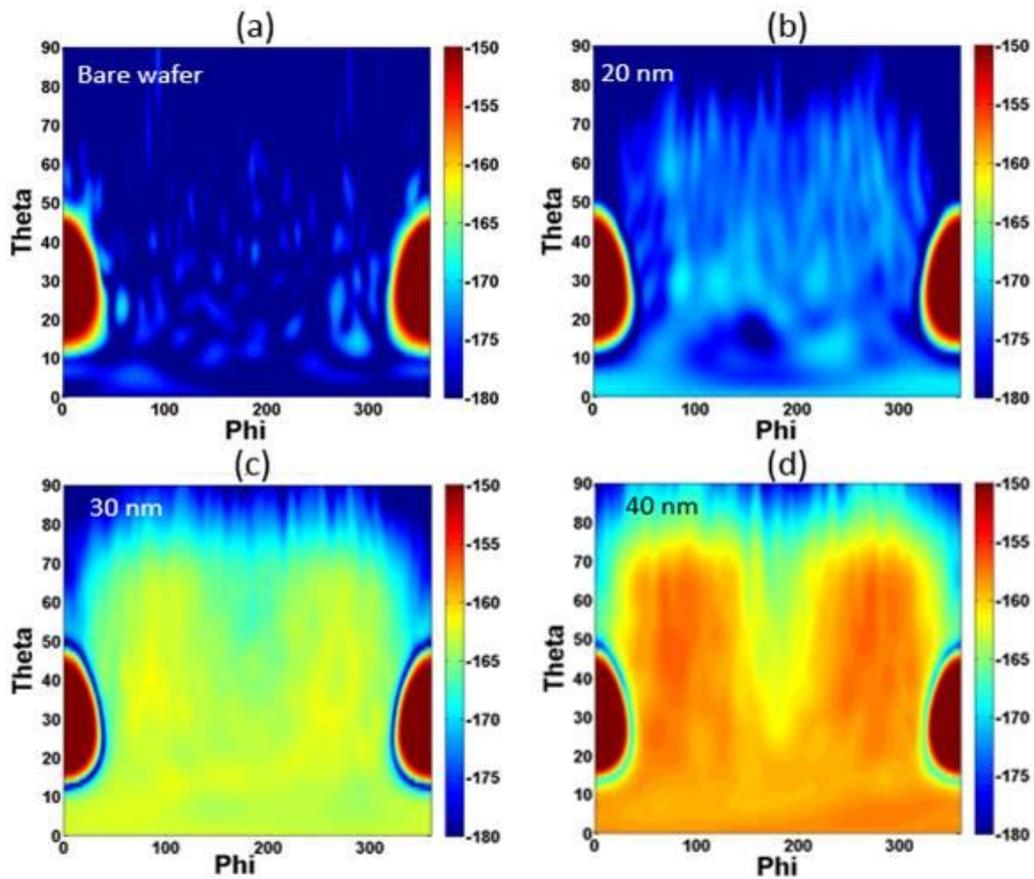

Figure 3. Light scattering by a) bare Si wafer, b) 20 nm sphere on a Si wafer, c) 30 nm sphere on a Si wafer and d) 40 nm sphere on a Si wafer, all of which are illuminated with a p-polarized Gaussian beam having a wavelength of 193 nm.

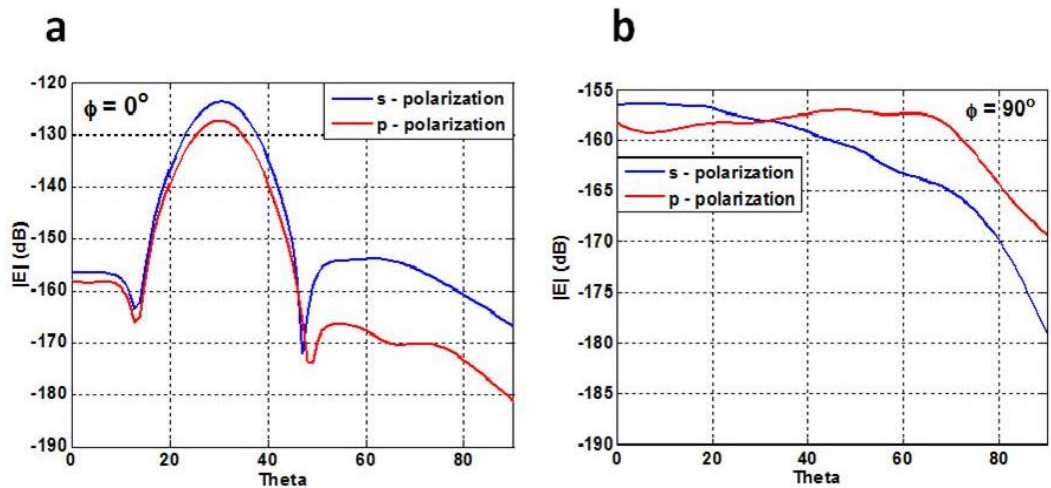

Figure 4. a) Scattering pattern ($\varphi = 0^0$) comparison for 40 nm radius sphere on wafer: s vs p polarization b) Scattering pattern ($\varphi = 90^0$) comparison for 40 nm radius sphere on wafer: s vs p polarization.

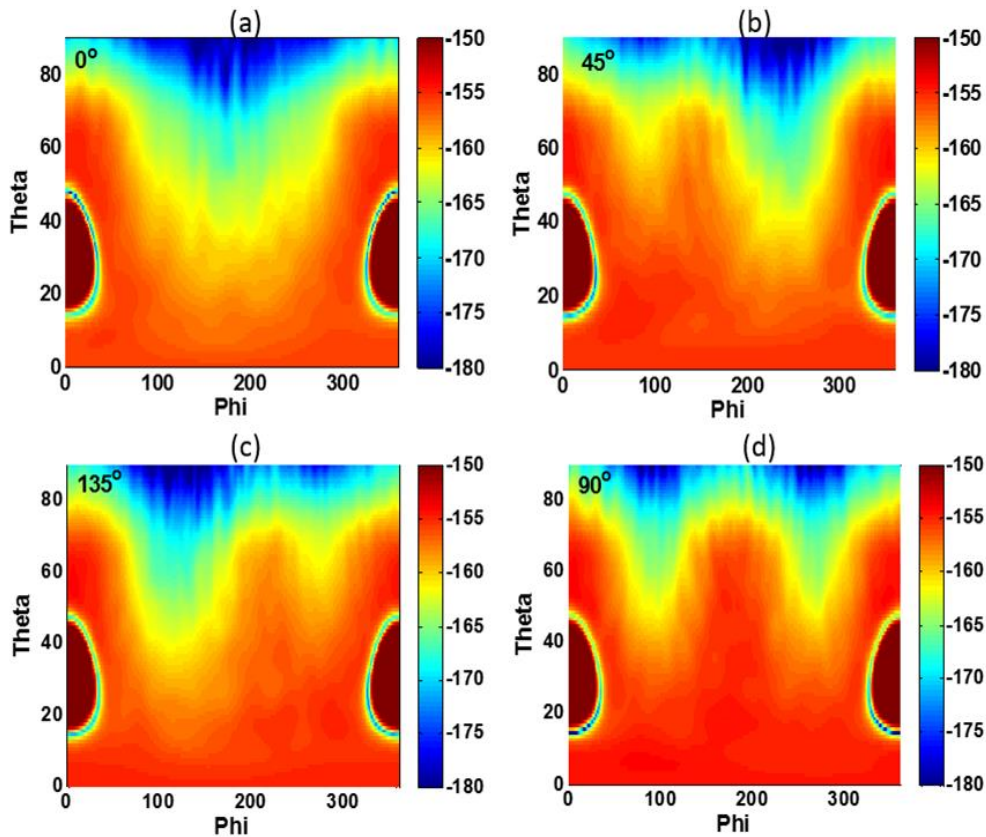

Figure 5. Light scattering by a) bare Si wafer, b) 20 nm sphere on a Si wafer, c) 30 nm sphere on a Si wafer and d) 40 nm sphere on a Si wafer, all of which are illuminated with a p-polarized Gaussian beam having a wavelength of 193 nm.

**Authors**

**Srikumar Sandeep** received his MS and PhD in electrical engineering from University of Colorado, Boulder in 2011 and 2012 respectively. From 2012 to 2013, he was a postdoctoral research associate at University of Colorado, Boulder. Presently, he is a postdoctoral researcher at Ecole Polytechnique, Montreal. He has more than 5 years of industrial experience in several fields of electrical engineering such as computational electromagnetics, signal integrity, embedded systems and software development. He is a co-inventor of 2 US patents. His technical interests include computational and applied electromagnetics.

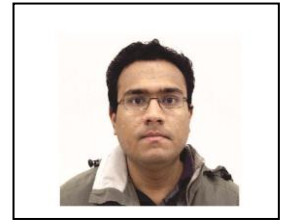

**Alexander Kokhanovsky** graduated from the Department of Physics at the Belarusian State University, Minsk, Belarus in 1983. He received his Ph.D in Optical Sciences from the B.I. Stepanov Institute of Physics, National Academy of Sciences of Belarus in 1991. His habilitation work (Main Geophysical Observatory, St.Petersburg, Russia, 2011) was devoted to the development of advanced cloud and snow remote sensing techniques based on spaceborne observations. His research interests include studies on light propagation and scattering in terrestrial atmosphere. Dr. Kokhanosky is the author of books *Light Scattering Media Optics: Problems and Solutions*, *Polarization Optics of Random Media*, *Cloud Optics* and *Aerosol Optics*. He has published more than 300 papers in the field of environmental optics, radiative transfer and light scattering

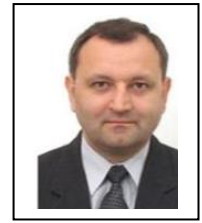